# Electronic structure and thermoelectric properties of CoTiSi *half*-Heusler alloy: Doping overtones

A. Shukla[a], Sadhana Matth[b], Raghavendra Pal[c], S.S.A. Warsi[a,*], Himanshu Pandey[b,#]

[a] Department of Physics, Integral University, Lucknow, Uttar Pradesh 226026, India
[b] Condensed Matter & Low-Dimensional Systems Laboratory, Department of Physics, Sardar Vallabhbhai National Institute of Technology, Surat 395007, India
[c] Department of Electronics Engineering, Sardar Vallabhbhai National Institute of Technology Surat-395007, Gujarat, India

Corresponding author e-mail: *salmanwarsi@iul.ac.in, #hp@phy.svnit.ac.in

**Abstract**

The quest for thermoelectric materials with high figures of merit is an ongoing and significant area of research. In this study, we investigate the thermoelectric properties of the CoTiSi *half*-Heusler alloy using density functional theory calculations implemented via the Wien2k package. Our approach begins with a thorough structural optimization to determine the equilibrium lattice parameter and the atomic positions of the constituent elements within the unit cell of CoTiSi. Following this, we analyze the thermal transport properties of the alloy under the constant relaxation time approximation, which allows us to gain insights into its thermoelectric performance. Our calculations reveal a substantial Seebeck voltage and thermopower, with notably higher values for *P-type* doping than *N-type* doping. This finding highlights the enhanced thermoelectric performance of *P-type* carriers in this material, providing a starting point for experimentalists to utilize this alloy for real device applications.

*Keywords: Heusler alloys; First principle Calculation, Thermoelectric applications*

1. **Introduction**

Research on Heusler alloys is rapidly expanding as researchers around the globe frequently propose and investigate new members within this family of alloys. Generally, these alloys are widely known for their applications in spintronics due to their half-metallic character. Moreover, researchers have experimentally and/or theoretically explored various newer properties and applications of these Heusler alloys, such as spin-caloritronics, anomalous Hall effect, spin Hall effect, topological insulator, Weyl semimetal, and catalytic activity [1-6]. Apart from these applications, an arena for clean and renewable energy generation that can generate electrical energy from heat, especially waste heat from industries, is also being extensively investigated [7-11]. This kind of energy conversion is given by a dimensionless parameter, known as figure-of-merit ($zT$), of a thermoelectric material.

$$zT = \frac{\sigma S^2}{\kappa} T \qquad (1)$$

Here, $S$ is the Seebeck coefficient, $\sigma$ is the electrical conductivity, $\kappa$ is the thermal conductivity with contributions from the electrons and phonons, and $T$ is the absolute temperature. Materials with a higher value of $zT$ provide more conversion efficiency. To maximize $zT$, a balance between the numerator $\sigma S^2$, also known as the power factor, and the denominator $\kappa$ is needed to maintain. Hence, semiconductor materials are considered the best for this electrical power generation [12–14]. Many efforts have been made to improve the value of $zT$ and achieve better energy conversion efficiency. In this direction, scattering through grain boundary [15], preferential phonon scattering [16], nanostructuring [17, 18], and solid-solution technique [19] *etc.* have been employed. Researchers have also tried to tailor the band structure by applying strain [20], creating resonant states [21], and optimum element doping [22]. With the application of strain, the electronic structure of a material can be changed and suitably tailored as per the desired requirements because the changes in the electronic structure are strongly correlated with the materials' characteristics, such as magnetic, optical, and electronic transport properties. Many strain-dependent investigations have recently been reported to improve materials' performance [23-32]. By utilizing density functional theory (DFT), a tetragonal strain is applied in the range of ±10%, and the half-metallic nature of *half*-Heusler (HH) alloy FeCrAs is maintained [32]. Yadav *et al.* investigated the consequences of uniform strain in the range of ±10% on the thermoelectrical and topological behaviour of BaPtS HH alloy. They reported a maximum value of $zT$ around 0.22 at room temperature under a strain of 1% [23]. In another work, via strain engineering, the $zT$ value is around 0.81 at room temperature, with feeble variation with temperature, as reported [24]. For the high-temperature application of thermoelectric materials such as HH alloy BiBaK, the $zT$ value is increased from 0.6 (unstrained case) to 0.93 under the application of an isotropic compressive strain of 9% (at 1200 K) [26]. An increase in $S$ and thermoelectric power of CoHfSi is also attained by applying both isotropic and tetragonal strains [33]. Since the Heusler compounds are very much prone to antisite disorder [1, 6], atomic swapping between the interpenetrated sub-lattices, stoichiometry imbalance, *etc.*, due to the use of nearby transition metal elements, a rigorous optimization of atomic positions is very much required before exploring any application-based characteristics. Also, investigation under the application of strain is worthwhile as when the actual devices are fabricated, they are usually under some strain (tensile/compressive) due to heteroepitaxy from the underneath layer or substrate. Hence, to correlate this, a study with strain can provide a direction to experimentalists for the better tuning of the materials' properties.

Being one of the most robust and mechanically stable alloys, Co-based HH alloys offer a tunable mid-to-high range of temperature applications. In this work, we first performed self-consistent electronic structural calculations for structure optimization of HH CoTiSi using the Wien2k *ab-initio* package. Many possible combinations of atomic sites for Co, Ti, and Si atoms and for an empty sub-lattice are considered. After that, uniform and tetragonal strains are applied to the most stable phase, and then the consequences of these strains on thermoelectric properties are examined.

2. **Computational Method**

We have performed the electronic band structural calculations on HH alloy CoTiSi using the Wien2k *ab-initio* package [34]. A generalized gradient approximation was utilized with a $K_{max}R_{MT}$ value of 9, where $K_{max}$ represents the plane wave cut-off and $R_{MT}$ is the radius of the muffin-tin sphere. To avoid overlapping the atomic spheres, $R_{MT}$ values for the Co, Ti, and Si elements are used as 2.21, 2.10, and 2.00, respectively. The crystal structure of these HH alloys consists of four interpenetrated sub-lattices with face-centred cubic (space group: 216) crystal structure, out of which one is empty for HH alloys. Three constituting atoms can occupy any three atomic positions out of these four in the unit cell: *A* (0, 0, 0), *B* (0.50, 0.50, 0.50), *C* (0.25, 0.25, 0.25), and *D* (0.75, 0.75, 0.75). So, various combinations of these atomic positions are possible for the HH alloys, depending on occupied atomic sites. Meticulous self-consistent calculations were performed on different possible combinations of atomic positions. A total of 24 combinations of atomic sites for Co, Ti, and Si atoms, as well as for the empty site within the unit cell, is considered for CoTiSi HH alloy. The calculated ground state energy and equilibrium lattice parameter, as estimated from the fitting by the Birch-Murnaghan equation of state for each combination, are summarized in Table 1. The lowest value of ground state energy is obtained for phase #9, confirming the most stable phase with Co atoms at (0.25, 0.25, 0.25), Ti atoms at (0, 0, 0), Si atoms at (0.50, 0.50, 0.50) and keeping (0.75, 0.75, 0.75) site empty. The corresponding optimized lattice parameter is estimated to be around 5.5843Å, comparable with the previously calculated and experimentally obtained value of 5.58 Å [9,11,35]. All further calculations were carried out for this phase #9. After investigating the phase stability of CoTiSi HH alloy, the effect of uniform and tetragonal strains on semiconducting bandgap and $zT$ is explored. This study will provide an insightful pathway for the experimentalist before synthesizing the thin films or related devices using CoTiSi for thermoelectric or other associated applications.

**Table 1:** List of all possible atomic sites for Co, Ti, and Si atoms, as well as the empty/vacancy site in the unit cell of CoTiSi HH alloy. Various atomic positions in the unit cell are given as A: (0, 0, 0), B: (0.50, 0.50, 0.50), C: (0.25, 0.25, 0.25) or D: (0.75, 0.75, 0.75). The last two columns provide the calculated ground state energy (in eV) and equilibrium lattice parameter (in Å).

| Phase | Co | Ti | Si | Vacancy | Total energy -5074 + …. (eV) | Equilibrium lattice parameter (Å) |
|---|---|---|---|---|---|---|
| 1 | A | B | C | D | -0.705032 | 5.6450 |
| 2 | A | B | D | C | -0.704782 | 5.6429 |
| 3 | A | C | B | D | -0.624047 | 5.7320 |
| 4 | A | D | B | C | -0.623762 | 5.7336 |
| 5 | A | C | D | B | -0.778256 | 5.5775 |
| 6 | A | D | C | B | -0.779503 | 5.5840 |
| 7 | B | A | C | D | -0.704904 | 5.6438 |
| 8 | B | A | D | C | -0.705588 | 5.6409 |
| 9 | C | A | B | D | -0.779509 | 5.5843 |
| 10 | D | A | B | C | -0.779503 | 5.5842 |
| 11 | D | A | C | B | -0.622373 | 5.7289 |
| 12 | C | A | D | B | -0.622372 | 5.7289 |
| 13 | B | C | A | D | -0.622314 | 5.7293 |
| 14 | B | D | A | C | -0.622370 | 5.7288 |
| 15 | C | B | A | D | -0.779499 | 5.5842 |
| 16 | D | B | A | C | -0.779498 | 5.5842 |
| 17 | C | D | A | B | -0.705618 | 5.6410 |
| 18 | D | C | A | B | -0.705614 | 5.6409 |
| 19 | B | C | D | A | -0.779499 | 5.5843 |
| 20 | B | D | C | A | -0.779499 | 5.5843 |
| 21 | C | B | D | A | -0.622364 | 5.7294 |
| 22 | D | B | C | A | -0.622369 | 5.7294 |

| 23 | C | D | B | A | -0.705618 | 5.6410 |
| 24 | D | C | B | A | -0.705589 | 5.6409 |

## 3. Result and discussion

### 3.1 Electronic Structure Properties

#### 3.1.1 Unstrained CoTiSi HH alloy

Figure 1 shows the plots for the total and atom-projected density of states (DOSs) for spin-up and spin-down bands, which reveal a distinct half-metallic nature with a gap of 1.1 eV. In the valence band (VB), the total DOSs for spin-up and spin-bands are dominated by the contribution from Co atoms. The contribution from another transition metal, *i.e.,* Ti atoms, is significantly small compared to Co atoms, with a lower energy range of less than -2.0 eV. But around the Fermi level ($E_F$), Co and Ti atoms contribute the most to the total DOS, along with a significant amount of contribution from the Si DOS. On the other hand, after the gap in the conduction band (CB), the Ti atom's DOSs are remarkably increased compared to those for Co atoms. Still, these two atoms mainly contribute to both bands' total DOSs for spin-up and spin-down. Here, the contribution from Si atoms is also feeble. Figure 2(a) depicts the band structure plot for the spin-up band, and the required semiconducting nature can also be seen from here. A similar kind of band structure plot is obtained for the spin-down band. An indirect bandgap (*L - X*) of around 1.18 eV is estimated with the occurrence of conduction band minimum (CBM) at 1.18 eV and valence band maximum (VBM) at -0.004 eV [as seen from Fig. 2(b)]. Valley degeneracies near $E_F$ can also be seen from the band structure plot; thus, a combination of heavier and lighter holes is suitable for TE performance. The comparatively flat band infers a larger effective mass, which increases *S* and, hence, *PF*. The sharp band implies holes with a lighter effective mass, contributing to highly mobile holes and increasing their conductivity [36-38]. This combination of heavy and light holes boosts the thermopower.

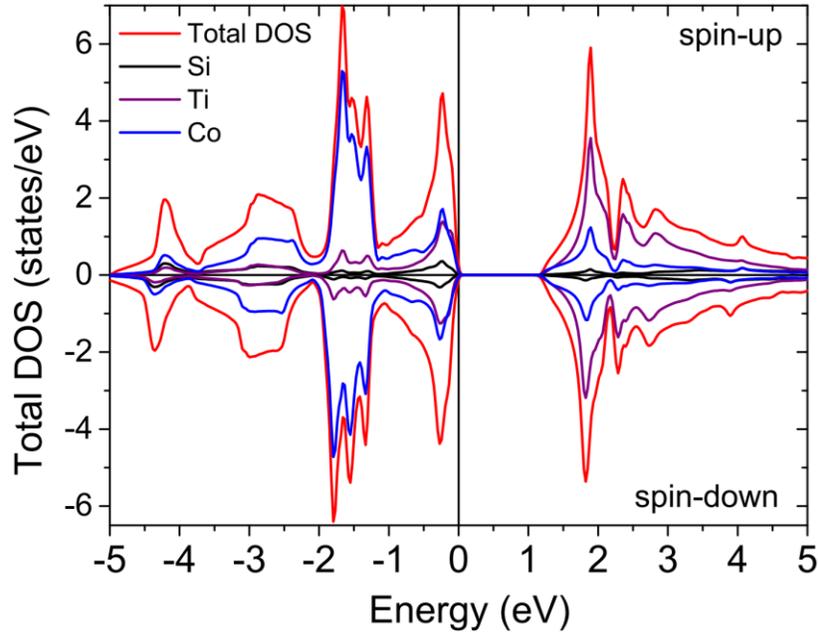

Figure 1: Total and atom-projected density of states plots for spin-up and spin-down bands of CoTiSi HH alloy.

### 3.1.2 Application of uniform strain

Firstly, we have applied the uniform (also known as isotropic) strain compressive and tensile up to 10% with respect to unstrained CoTiSi unit cell with a lattice parameter of 5.5843 Å. Here, all the lattice parameters ($a$, $b$, and $c$) are changed uniformly in all three directions, and hence, a change in the unit cell volume also occurs. Under uniform compressive strain, an isotropic pressure is applied to the unit cell to reduce all the lattice parameters by the same amount. In contrast, in the case of uniform tensile strain, the unit cell is expanded isotopically by increasing all the lattice parameters by the same amount.

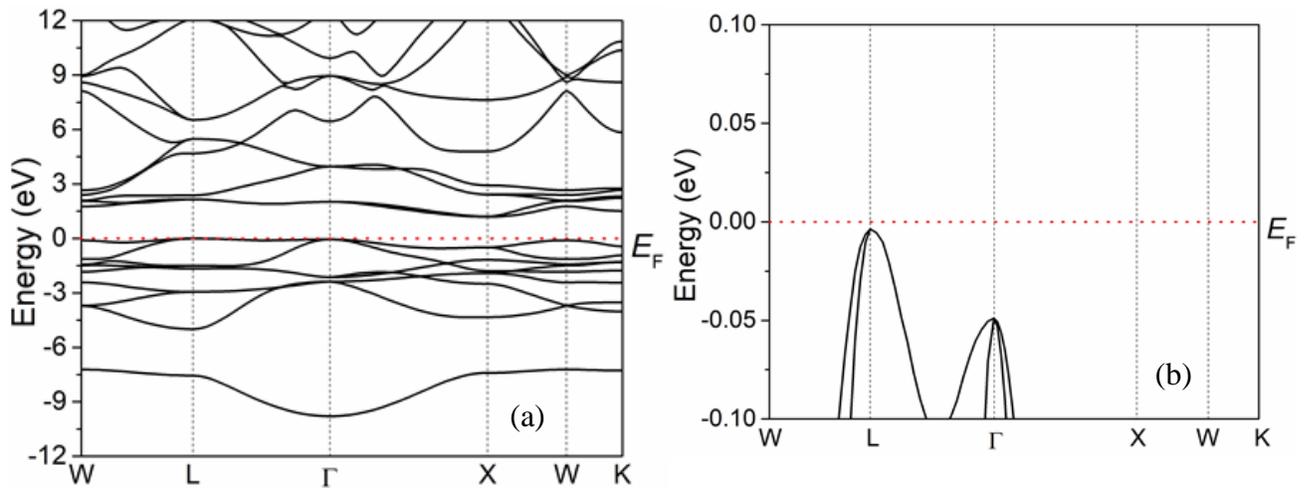

**Figure 2**: Band structure plot spin-up band of CoTiSi HH alloy, (b) For a closer look of valence band maxima, a zoomed version of (a) in the energy range of ± 0.10 eV

**Table 2**: Lattice parameter (*a*), valence band maximum and conduction band minimum position, and bandgap under uniform strain applied to CoTiSi unit cell.

| Strain (%) | *a* (Å) | VBM (eV) | CBM (eV) | band gap (eV) |
|---|---|---|---|---|
| -10 | 5.0259 | - | - | Metal |
| -9 | 5.0817 | - | - | Metal |
| -8 | 5.1376 | - | - | Metal |
| -7 | 5.1934 | - | - | Metal |
| -6 | 5.2492 | - | - | Metal |
| -5 | 5.3051 | 0.09 | 1.30 | 1.21 |
| -4 | 5.3609 | 0.06 | 1.26 | 1.20 |
| -3 | 5.4168 | 0.04 | 1.26 | 1.22 |
| -2 | 5.4726 | 0.01 | 1.25 | 1.24 |
| -1 | 5.5285 | 0.009 | 1.22 | 1.21 |
| -0.5 | 5.5564 | -0.002 | 1.20 | 1.20 |
| 0 | 5.5843 | -0.004 | 1.18 | 1.18 |
| 0.5 | 5.6122 | -0.002 | 1.17 | 1.17 |
| 1 | 5.6401 | -0.001 | 1.10 | 1.10 |
| 2 | 5.6960 | -0.03 | 1.12 | 1.15 |
| 3 | 5.7518 | -0.01 | 1.06 | 1.07 |
| 4 | 5.8077 | 0.02 | 1.01 | 0.99 |
| 5 | 5.8635 | 0.06 | 0.99 | 0.93 |
| 6 | 5.9194 | 0.05 | 0.91 | 0.86 |
| 7 | 5.9752 | 0.01 | 0.90 | 0.89 |
| 8 | 6.0310 | 0.01 | 0.88 | 0.87 |
| 9 | 6.0869 | - | - | Metal |
| 10 | 6.1427 | - | - | Metal |

Figure 3(a) shows the variation of total energy and bandgap with applied uniform strain. As deviating from the unstrained case, the total energy increases from its minimum, which corresponds to the equilibrium lattice parameter. The bandgap gradually changes and remains in the 0.87 – 1.24 eV range. The estimated values of the equilibrium lattice parameter, the location of VBM and CBM, and the bandgap under the tensile and compressive uniform strains are summarized in Table 1. By changing the strain, CBM changes almost monotonically from 1.30 eV for -5 % to 0.88 eV for +8 %, whereas VBM remains close to either side of $E_F$. A maximum bandgap of 1.24 eV is estimated for -2% strain. In some cases, VBM crosses $E_F$ by a minimal value of energies and DOSs. This can be understood as a *p*-type semiconductor as very few (< 10 meV) states are available at $E_F$, which may be due to an increase in electron energy with strain that causes electrons to cross $E_F$. Beyond 5% compressive uniform strain and 8% tensile uniform strain, metallic behaviour is established in this alloy.

### 3.1.3 Application of tetragonal strain

Under the application of tetragonal strain, the volume of the unit cell is considered to be constant. If the *c*-axis lattice parameter is elongated, then the other two sides, *a* and *b*, will be reduced in such a way that the volume of the unit cell remains the same as for the unstrained unit cell. An isotropic strain is assumed for in-plane lattice parameters *a* and *b*. Under the compressive tetragonal strain, the *c*-axis lattice parameter is decreased (or *a* is increased) with respect to unstrained value, whereas for the case of tensile tetragonal strain, the *c*-axis lattice parameter is increased (or *a* is decreased).

**Table 3**: Lattice parameter (*a*), valence band maximum and conduction band minimum position, and bandgap under tetragonal strain applied to CoTiSi unit cell.

| Strain (%) | *c* (Å) | *a* (Å) | VBM (eV) | CBM (eV) | band gap (eV) |
|---|---|---|---|---|---|
| -4 | 5.3609 | 5.6995 | - | - | metal |
| -3 | 5.4168 | 5.6700 | 0.07 | 1.09 | 1.02 |
| -2 | 5.4726 | 5.6410 | 0.02 | 1.16 | 1.14 |
| -1 | 5.5285 | 5.6124 | 0.007 | 1.20 | 1.19 |
| -0.5 | 5.5564 | 5.5983 | 0.002 | 1.20 | 1.20 |
| 0 | 5.5843 | 5.5843 | -0.004 | 1.18 | 1.18 |
| 0.5 | 5.6122 | 5.5704 | 0.004 | 1.19 | 1.19 |
| 1 | 5.6401 | 5.5564 | 0.009 | 1.17 | 1.16 |
| 2 | 5.6960 | 5.5293 | 0.007 | 1.16 | 1.15 |
| 3 | 5.7518 | 5.5024 | 0.01 | 1.12 | 1.11 |
| 4 | 5.8077 | 5.4759 | - | - | metal |

Figure 3(b) depicts the variation of total ground state energy and bandgap with applied tetragonal strain. The total energy increases as we move away from either side of the equilibrium. This is due to strain energy in the unit cell. The bandgap values are almost similar for each case, ranging from 1.02 – 1.20 eV. A maximum bandgap of 1.20 eV is estimated for a compressive tetragonal strain of 0.5%. The estimated values of the equilibrium lattice parameter, the location of VBM and CBM, and the bandgap under the tensile and compressive uniform strains are summarized in Table 2. Here, beyond 3%, a metallic nature is also established in this alloy. Except for the unstrained case, VBM slightly crosses $E_F$. The number of such states available in VB above $E_F$ increases marginally with the tetragonal strain of either type. For the limiting case of ± 4%, a metallic nature is observed. Hence, since the electronic structure of CoTiSi is altered by strain, which ultimately changes the DOS at $E_F$, its thermal transport properties can also be modified suitably.

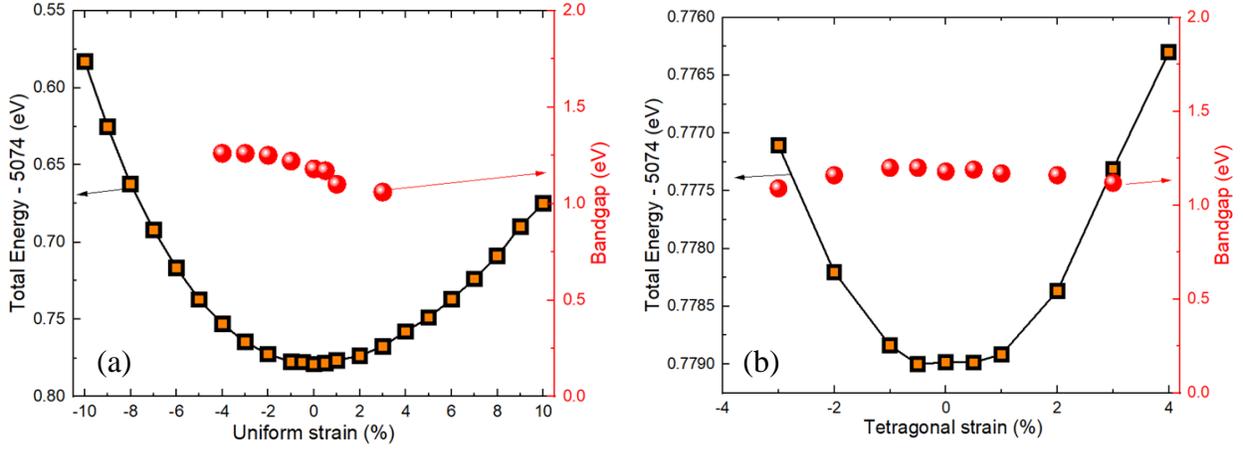

**Figure 3**: The variation of total energy as estimated from self-consistent calculation (left Y-axis, black colour) and bandgap (right Y-axis, red colour) for CoTiSi under (a) uniform strain and (b) tetragonal strain.

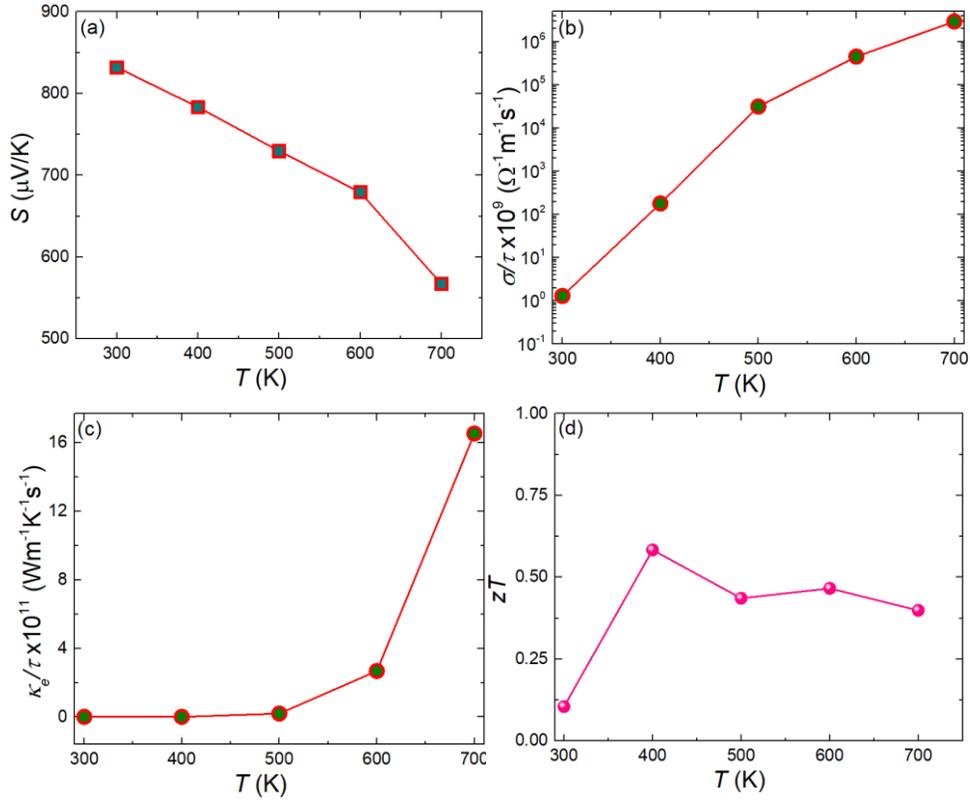

Figure 4: Variation of (a) $S$, (b) $\sigma/\tau$, (c) $\kappa/\tau$, and (d) $zT$ with temperature for unstrained CoTiSi *half*-Heusler alloy.

### 3.2 Transport properties

#### 3.2.1 Without strain

Various transport properties are calculated using the BoltzTraP2 package [39], by which semiclassical Boltzmann transport equations are solved. A rigid band approximation and constant relaxation time approximation (CRTA) are assumed while solving those equations. For the former case, an assumption that the electronic band structure does not change significantly due to any variation of doping and temperature is considered, whereas for the latter case, a constant relaxation time ($\tau$) is assumed throughout the Brillouin zone. It does not change with the band energies due to doping and temperature. A comparatively dense k-mesh of 40 x 40 x 40 is used for electronic band structural calculations to calculate the transport coefficients, as they depend on the derivative of electrons' band energy with respect to the k-point. Figure 4 (a-d) shows the temperature-dependent variation of $S$, $\sigma/\tau$, $\kappa/\tau$, and $zT$, respectively, in the temperature range 300–700 K. With the increase in temperature, $S$ decreases whereas the quantities $\sigma/\tau$ and $\kappa/\tau$ increase. Under CRTA, a good semiconducting feature is also established from Fig. 4 (b and c) and by using Eq. (1), the estimated values of $zT$ are plotted in Fig. 4(d).

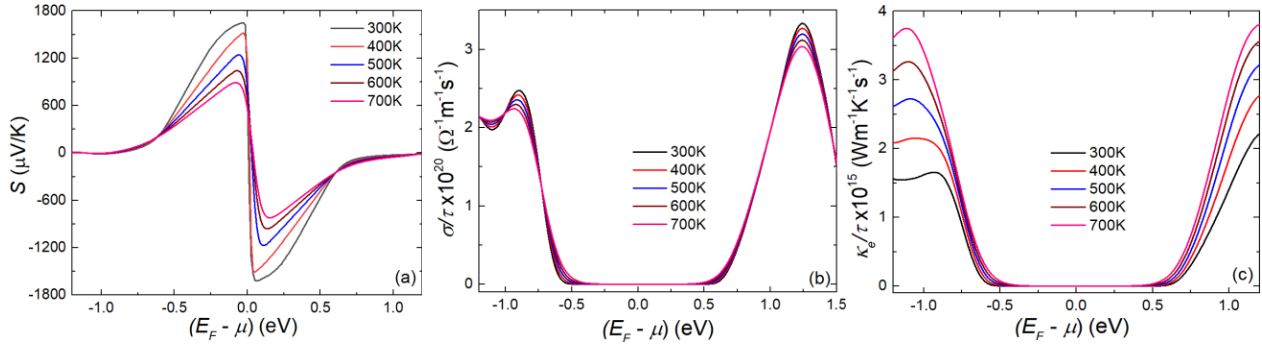

Figure 5: Variation of (a) $S$, (b) $\sigma/\tau$, and (c) $\kappa/\tau$ with the position of the Fermi level temperature for unstrained CoTiSi half-Heusler alloy. Here, $\mu$ is the Fermi energy at zero temperature.

The variation of thermoelectric properties such as $S$, $\sigma/\tau$, and $\kappa/\tau$ with the position of the Fermi level ($E_F - \mu$) are depicted in Fig. 5 (a-c), respectively for unstrained CoTiSi HH alloy. Here, $\mu$ is the Fermi energy defined for absolute zero temperature. These properties are calculated for different temperatures from 300 K to 700 K. The values with $E_F < \mu$ and $E_F > \mu$ represent the P- and N-type behaviour. Assuming the rigid band approximation, where the original features of the band structure do not change with influencing parameters such as temperature, strain, doping concentration, *etc.*, the modified position of the Fermi level represents the effect of those external parameters. From Fig. 5 (a), one can infer that the values of $S$ are positive for $E_F = \mu$, suggesting a P-type behaviour of unstrained CoTiSi at the ground state. This agrees with the earlier observation, as shown in Fig. 4 (a). This is mainly due to the position of the VBM, which is just below (around 2 meV) the Fermi level, and any increase in temperature lowers the Fermi level. Also, $S$ can attain the maximum for some selective doping cases as we move on either side of $\mu = E_F$. The maximum value of $S$ decreases rapidly with the increase in temperature due to an increase in carrier concentration with temperature. At 300 K, a maximum value of $S$ is found to be around 1645 $\mu$V/K. The $\sigma/\tau$ and $\kappa/\tau$ values are near their minimum around $E_F = \mu$ and up to $E_F - \mu = \pm 0.4$ eV. Very feeble dependence on $\kappa/\tau$ with the position of the Fermi level, is observed, whereas $\sigma/\tau$ is found almost constant up to the range $E_F - \mu = \pm 0.4$ eV. So, to have a high value of $zT$, the optimized combination of $S$, $\sigma/\tau$, and $\kappa/\tau$ should be taken rather than choosing a maximum of $S$ and a minimum of $\kappa/\tau$. Hence, by controlling the degree of strain and carrier concentration, the doping levels can be optimized, and further, the performance of a TE device can be tailored. Suitable doping will access the utilization of the same alloy as P-type or N-type.

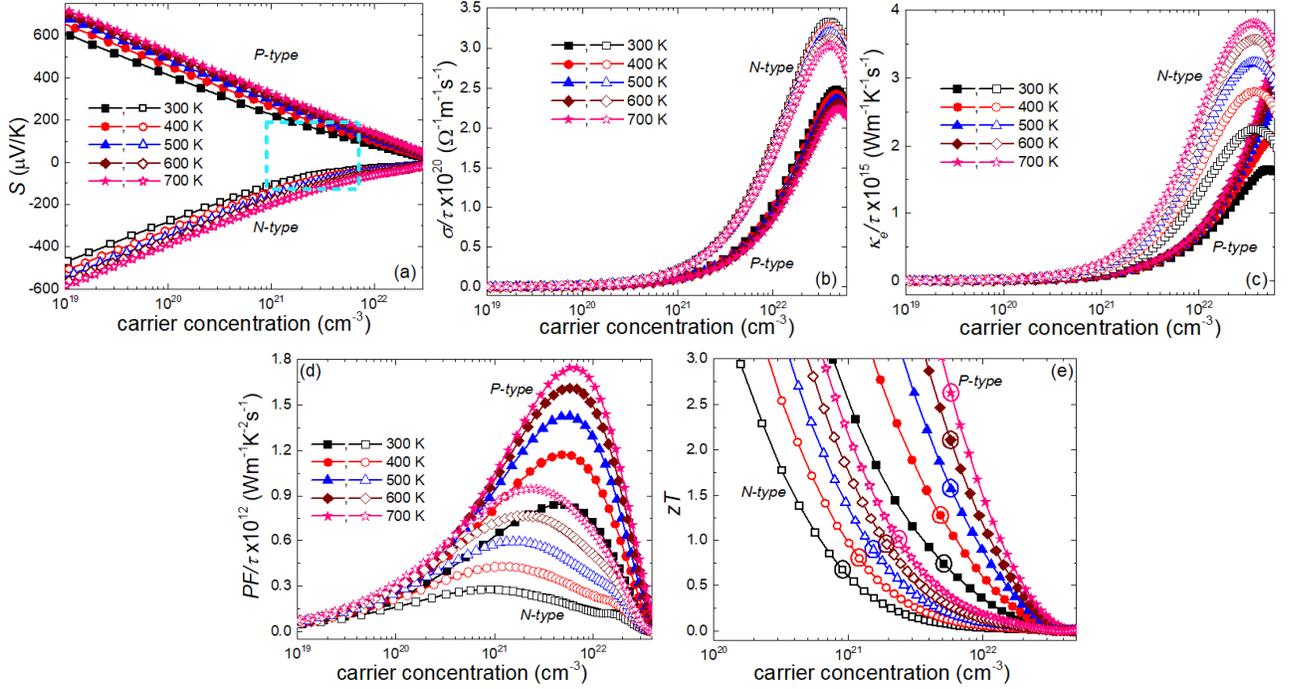

Figure 6: Variation of (a) $S$, (b) $\sigma/\tau$, (c) $\kappa/\tau$, (d) $PF/\tau$, and (e) $zT$ with carrier concentration for different temperatures for CoTiSi *half*-Heusler alloy. Here, the open and closed symbols are used for *N-type* and *P-type* to semiconductors. For more details, please refer to text.

**Table 4:** Optimal doping concentration (x $10^{21}$ cm$^{-3}$), doping levels (carrier/unit cell) and corresponding values for Seebeck coefficient (μV/K), maximum thermoelectric power factors with respect to relaxation time $\tau$ ($S^2\sigma/\tau$, $10^{12}$ W m$^{-1}$K$^{-2}$s$^{-1}$), and $zT$ estimated at maximum power factor for both *P*- and *N-type* doping levels.

| T (K) | P-type | | | | | N-type | | | | |
|---|---|---|---|---|---|---|---|---|---|---|
| | Carrier conc. | doping level | S | $PF_{max}/\tau$ | $zT$ at $PF_{max}$ | Carrier conc. | doping level | S | $PF_{max}/\tau$ | $zT$ at $PF_{max}$ |
| 300 | 5.13 | 0.223 | 121 | 0.84 | 0.74 | 0.91 | 0.040 | -120 | 0.28 | 0.67 |
| 400 | 4.85 | 0.211 | 150 | 1.17 | 1.28 | 1.21 | 0.053 | -129 | 0.43 | 0.80 |
| 500 | 5.82 | 0.254 | 157 | 1.43 | 1.57 | 1.55 | 0.067 | -135 | 0.59 | 0.89 |
| 600 | 5.78 | 0.251 | 172 | 1.62 | 2.11 | 1.93 | 0.084 | -138 | 0.77 | 0.96 |
| 700 | 5.83 | 0.254 | 183 | 1.75 | 2.63 | 2.37 | 0.103 | -140 | 0.95 | 1.01 |

Achieving a high value of *zT* is crucial for enhancing thermoelectric efficiency; however, optimizing *PF* is equally important. To maximize the *PF* values, it is essential to understand the optimal doping level, which serves as a valuable reference point for experimentalists aiming to narrow down the doping window. The degree of doping can be characterized by the concentration of charge carriers and their adequate numbers, which ultimately dictate the semiconducting behaviour of either *P*- or *N-type*. In our analysis, we provide plots of *S*, $\sigma/\tau$, $\kappa/\tau$, $PF/\tau$, and *zT* as functions of carrier concentration for both *P*- and *N-type* CoTiSi, as illustrated in Fig. 6 (a-e). Notably, we observe that *S* reaches its peak around zero doping, as shown in Fig. 5(a), followed by a nearly monotonic increase with increasing carrier concentration for both doping types. Our estimated values of *S* are comparable to those of PbTe, a well-established thermoelectric material [40]. A similar trend is observed for the parameters $\sigma/\tau$ and $\kappa/\tau$, where a significant dependence emerges beyond a concentration of $10^{21}$ cm$^{-3}$. Figure 6(d) depicts the variation of $PF/\tau$ (defined as $S^2\sigma/\tau$) with carrier concentration. For both *P*- and *N-type* CoTiSi, a peak is observed for each plot of *PF/τ* taken for different temperatures. Although *PF/τ* is primarily influenced by *S* (due to its square dependence), the peak in *PF/τ* aligns closely with the peak for *σ/τ*. This correlation suggests a higher DOS at the Fermi level, delivering the maximum power factor (*PF$_{max}$*). Figure 6(e) presents the carrier concentration-dependent *zT* calculated at various temperatures, with specific data points highlighted to indicate *zT* values obtained at *PF$_{max}$/τ*. These observations allow us to estimate the optimal doping levels for synthesizing *P-* and/or *N-type* CoTiSi based on specific application needs. To achieve substantial thermopower, a material must possess a higher DOS at its band edges [41, 42]. A large DOS at VBM or CBM indicates that the material will exhibit a high *PF* for *P-* or *N-type* doping, respectively. In the case of CoTiSi, our analysis reveals a significantly higher DOS at the VBM than the CBM, supporting our inference that CoTiSi will likely provide a more significant *PF* for *P-type* doping. Under the constant relaxation time approximation, our estimated values of *PF$_{max}$/τ*, presented in Fig. 7 and summarized in Table 4, reinforce this conclusion. Assuming $τ = 10^{-14}$ s, we estimate *PF$_{max}$/τ* at 300 K to be 84 x $10^{-4}$ Wm$^{-1}$K$^{-2}$ for *P-type* CoTiSi and 28 x $10^{-4}$ Wm$^{-1}$K$^{-2}$ for *N-type* CoTiSi. Compared to PbTe [43], which has a thermopower of 35 x $10^{-4}$ Wm$^{-1}$K$^{-2}$, our results for *P-type* CoTiSi are significantly higher, while the *N-type* value remains slightly lower. The optimal carrier concentration range for achieving *PF$_{max}$/τ* is highlighted by a dashed rectangle in Fig. 6(a). Within the doping concentration range of $10^{21} – 10^{22}$ cm$^{-3}$, we find elevated values of *PF*. At 300 K, the Fermi level for *P-* and *N-type* CoTiSi is approximately 0.67 eV below the VBM and above the CBM, respectively. Additionally, the non-parabolic nature of VB near the Fermi level, particularly along the *W-L* and *L-Γ* directions, contributes to the high *PF* observed for *P-type* doping.

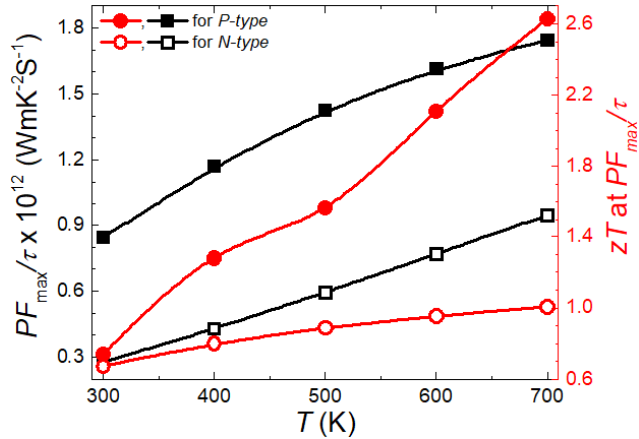

**Figure 7:** The temperature-dependence of maximum thermopower *PF$_{max}$/τ* and *zT* value at *PF$_{max}$*, for both *P-type* and *N-type* CoTiSi.

**Conclusion**

We have conducted first-principles calculations to optimize the structural phase of the CoTiSi *half*-Heusler alloy, a promising candidate for thermoelectric applications. Initially, we focused on refining the atomic positions of the constituent atoms within the unit cell. Through comprehensive unit cell relaxation and ground state energy minimization, we determined the equilibrium lattice parameter corresponding to the most stable phase of the alloy. Our calculations revealed an indirect bandgap of 1.18 eV, indicating potential for electronic applications. Notably, we observed significant Seebeck voltage and thermopower values that are competitive with those of established thermoelectric materials, such as PbTe. The large density of states near the Fermi level and its degeneracy and rapid variation at the valence band maximum were identified as critical factors contributing to the enhanced power factor for *P-type* doped CoTiSi. Moreover, we found that the thermopower exhibits a pronounced peak as a function of carrier concentration, suggesting a strong dependence on doping levels. Our analysis enabled us to identify optimal doping levels corresponding to maximum thermopower for both *P-type* and *N-type* scenarios. These theoretical insights underscore the unique properties of CoTiSi and point to its potential as a viable thermoelectric material. In summary, our findings provide comprehensive insights into the thermoelectric properties of CoTiSi, highlighting its potential for optimization through targeted doping strategies. This knowledge will facilitate the development of high-performance thermoelectric materials for practical applications.

**Conflicts of interest**

The authors declare that there are no conflicts of interest.

**Acknowledgement**

HP acknowledges the Science and Engineering Research Board (SERB), Govt. of India, for the research grant against the scheme ECR/2017/001612 and SVNIT seed research grant 2021-22/DOP/04. AS and SSAW are thankful to the Research and Development (R&D) wing of Integral University, Lucknow, for providing the support and manuscript communication number (IU/R&D2022–MCN0000xxxx) for the publication of this manuscript.